\documentclass[a4paper,conference]{IEEEtran}
\IEEEoverridecommandlockouts
\usepackage{cite}
\usepackage{amsmath,amssymb,amsfonts}
\usepackage{algorithmic}
\usepackage{graphicx}
\usepackage{listings}
\usepackage{textcomp}
\usepackage{xcolor}
\usepackage{balance}
\usepackage{booktabs}
\usepackage{tikz}
\usepackage{pgfplots}
\usepackage{tcolorbox}
\usepackage{adjustbox}
\usepackage{tabularx}
\usepackage{threeparttable}
\usepackage{multirow}
\usepackage{url}
\usepackage{hyperref}
\usepackage{colortbl}
\usepackage{pgf-pie}
\usepackage[capitalize,nameinlink,noabbrev]{cleveref}
\crefname{figure}{Fig.}{Figs.}
\usepackage[left=1.65cm,right=1.65cm,top=1.9cm]{geometry}

\usepgfplotslibrary{statistics}

\def\BibTeX{{\rm B\kern-.05em{\sc i\kern-.025em b}\kern-.08em
    T\kern-.1667em\lower.7ex\hbox{E}\kern-.125emX}}

\begin{document}

\newcommand{\RqOne}{RQ$_1$: What are the most frequently discussed keywords related to React on Stack Overflow?}

\newcommand{\RqTwo}{RQ$_2$: What types of React-related errors are commonly inquired about by users on Stack Overflow?}

\newcommand{\RqThree}{RQ$_3$: How do React-related errors vary for different user reputations?}

% \title{An Empirical Study of React-Library Related Issues Shared on Stack Overflow}
\title{Hot Topics and Common Challenges: an Empirical Study of React Discussions on Stack Overflow}
% {\footnotesize \textsuperscript{*}Note: Sub-titles are not captured for https://ieeexplore.ieee.org  and
% should not be used}
% \thanks{Identify applicable funding agency here. If none, delete this.}

% \author{Anonymous Author(s)} %double blind

% \makeatletter
% \newcommand{\linebreakand}{%
%   \end{@IEEEauthorhalign}
%   \hfill\mbox{}\par
%   \mbox{}\hfill\begin{@IEEEauthorhalign}
% }
% \makeatother

\makeatletter
\author{
\begin{@IEEEauthorhalign}
\IEEEauthorblockN{\hspace*{-0.05cm}1\textsuperscript{st} Yusuf Sulistyo Nugroho}
\IEEEauthorblockA{\hspace*{-0.05cm}\textit{Informatics Engineering} \\
\hspace*{-0.05cm}\textit{Universitas Muhammadiyah Surakarta}\\
\hspace*{-0.05cm}Surakarta, Indonesia \\
\hspace*{-0.05cm}yusuf.nugroho@ums.ac.id}
\\ [1.5ex]
\IEEEauthorblockN{\hspace*{-0.05cm}4\textsuperscript{th} Mohammed Humayun Kabir}
\IEEEauthorblockA{\hspace*{-0.05cm}\textit{Comp. Sci. and Telecommunication Eng.} \\
\hspace*{-0.05cm}\textit{Noakhali Science and Technology Univ.}\\
\hspace*{-0.05cm}Noakhali, Bangladesh \\
\hspace*{-0.05cm}kabir@nstu.edu.bd}
\and
\IEEEauthorblockN{2\textsuperscript{nd} Ganno Tribuana Kurniaji}
\IEEEauthorblockA{\textit{Electrical Eng. and Information Tech.} \\
\textit{Universitas Gadjah Mada}\\
Yogyakarta, Indonesia \\
gannotribuanakurniaji@mail.ugm.ac.id
}
% \and
\\ [1.5ex]
\IEEEauthorblockN{5\textsuperscript{th} Vanesya Aura Ardity}
\IEEEauthorblockA{\textit{Informatics Engineering}\\
\textit{Universitas Muhammadiyah Surakarta}\\
Surakarta, Indonesia\\
L200210170@student.ums.ac.id}
\and
\IEEEauthorblockN{\hspace*{0.05cm}3\textsuperscript{rd} Syful Islam}
\IEEEauthorblockA{\hspace*{0.05cm}\textit{Computer Science and Engineering} \\
\hspace*{0.05cm}\textit{Gopalganj Science and Technology Univ.}\\
\hspace*{0.05cm}Gopalganj, Bangladesh \\
\hspace*{0.05cm}syfulcse@gstu.edu.bd} 
\\ [1.5ex]
\IEEEauthorblockN{\hspace*{0.05cm}6\textsuperscript{th} Md. Kamal Uddin}
\IEEEauthorblockA{\hspace*{0.05cm}\textit{Comp. Sci. and Telecommunication Eng.} \\
\hspace*{0.05cm}\textit{Noakhali Science and Technology Univ.}\\
\hspace*{0.05cm}Noakhali, Bangladesh \\
\hspace*{0.05cm}kamaluddin@nstu.edu.bd}
\end{@IEEEauthorhalign}
}
\makeatother

\maketitle

\begin{abstract}
React is a JavaScript library used to build user interfaces for single-page applications. Although recent studies have shown the popularity and advantages of React in web development, the specific challenges users face remain unknown. 
Thus, this study aims to analyse the React-related questions shared on Stack Overflow. 
The study utilizes an exploratory data analysis to investigate the most frequently discussed keywords, error classification, and user reputation-based errors, which is the novelty of this work. 
The results show the top eight most frequently used keywords on React-related questions, namely, code, link, vir, href, connect, azure, windows, and website. 
The error classification of questions from the sample shows that algorithmic error is the most frequent issue faced by all groups of users, where mid-reputation users contribute the most, accounting for 55.77\%. 
This suggests the need for the community to provide guidance materials in solving algorithm-related problems.
We expect that the results of this study will provide valuable insight into future research to support the React community during the early stages of implementation, facilitating their ability to effectively overcome challenges to adoption.
\end{abstract}

\begin{IEEEkeywords}
error types, keywords, react, stack overflow, user reputation
\end{IEEEkeywords}

\section{Introduction}

React is a popular open-source JavaScript library applied for the custom user interfaces in single-page applications development, facilitating developers to build responsive web applications capable of data modification without reloading the page~\cite{keshari2023web}. 
The main objectives of React are to make web development faster, easier to scale, and simpler~\cite{komperla2022react}. 
However, despite its advantages, developers still face challenges when implementing React into practice. 
These challenges have made developers to find solutions by posting and discussing their problems on online question-and-answer (Q\&A) platforms like Stack Overflow (SO)~\cite{bangash2019developers, wu2019developers, zhou2020bounties}, a popular Q\&A website that serves as a repository of technical knowledge, promoting problem-solving and communication within the developer community~\cite{liu2021characterizing, an2017stack}.

A number of studies have focused on the quality and content of SO, such as user topics and challenges~\cite{islam2021network}, comment evolution~\cite{zhang2019reading}, machine learning~\cite{ahmad2020systematic}, chatbot~\cite{abdellatif2020challenges}, and privacy~\cite{tahaei2020understanding}.
% For example, Islam et al.~\cite{islam2021network} investigated questions on SO related to the network simulator to gain insights regarding user topics and the challenges they encounter. 
% Prior study~\cite{zhang2019reading} analyzed user discussions within comments on SO, with a specific focus on the changes in comments. 
% Similarly, a study by Ahmad et al.~\cite{ahmad2020systematic} aimed to systematically identify and classify Machine Learning algorithms used for software requirements on the platform, 
% Abdellatif et al.~\cite{abdellatif2020challenges}, on the other hand, used topic modelling techniques to determine the popularity and complexity of chatbot-related topics on the platform, 
% while Tahaei et al.~\cite{tahaei2020understanding} clarify the challenges and difficulties when navigating privacy-related topics.
% 
% 
In addition to these recent questions, other studies have addressed various aspects of SO and its user community. 
% For example, Islam et al.~\cite{islam2021network} investigated question posts on SO related to the network simulator to gain insights regarding user topics and the challenges they encounter. 
% Furthermore, a prior study~\cite{zhang2019reading} conducted an in-depth analysis of user discussions within comments on SO, with a specific focus on the changes in comments. 
Beyer et al.~\cite{beyer2020kind} focused on automating the classification of question posts on the platform, while May et al.~\cite{may2019gender} investigated how gender variations in SO community behaviour patterns and outcomes. 
Another prior work analysed the contribution of comments posted on SO~\cite{zhang2019reading}.
In addition, Lopez et al.~\cite{lopez2019anatomy} conducted a survey to investigate the prevalence and support for secure coding practices on the platform. 
In the context of software development and knowledge sharing, these works contribute a substantial body of research that attempts to understand the nature, challenges, and implications of SO.

\begin{figure*}
    \centering
    \includegraphics[width=.85\textwidth]{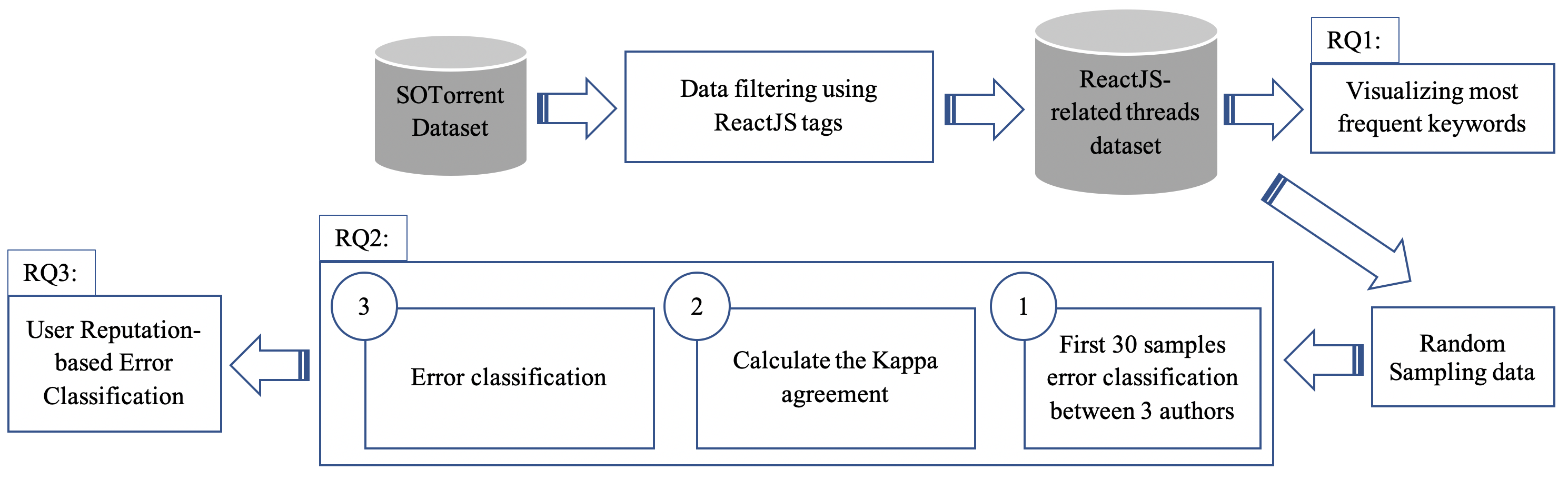}
    \caption{Methodology of the data collection and analysis processes in this study.}
    \label{fig:studysteps}
\end{figure*}

Despite the increasing popularity of the React library topics and its adoption in web development~\cite{islam2022exploration, kurniaji2023preliminary},
% as seen by recent studies and the discussion on SO~\cite{islam2022exploration, kurniaji2023preliminary}, 
developers frequently encounter issues related to algorithmic logic, state management, and debugging. 
% While prior studies~\cite{terzi2022software, wang2023empirical, torres2023investigation} have explored general JavaScript-related challenges, a comprehensive analysis specific to React issues on SO remains limited. 
% Therefore, this study aims to fill this gap by investigating the types of errors that relate to React from three perspectives (i.e., frequently discussed keywords, error classification, and their correlation with user expertise) shared on SO, which is the novelty of this work.
Although JavaScript-related issues have been studied~\cite{terzi2022software, wang2023empirical, torres2023investigation}, specific challenges in React remain underexplored.
Thus, this study aims to address this gap, which is the novelty of this work.
The insights gained will benefit both novice developers and documentation providers in structuring more effective learning materials.
Using the SOTorrent dataset, a publicly accessible resource based on the official SO data dump~\cite{baltes2019sotorrent}, this study offers a new perspective.

In this paper, we applied an exploratory study~\cite{sahoo2019exploratory} to investigate the most frequently discussed keywords, the error classification, and the error classifications based on user reputations. 
The dataset used in this study was extracted from the publicly available SOTorrent~\cite{baltes2019sotorrent} based on specified tags related to React, resulting in 447,542 questions out of 22,087,195 question posts. 
To manually classify problem-based question types, a sample of 384 React-related question posts is utilized.

The results show that the eight most frequent React-related keywords in SO question posts can serve as a resource for React developers seeking solutions and designing better tutorials for React users for specific problems. 
In addition, we observed that React users mostly face algorithmic errors, making them the dominant issue on SO.
Finally, all groups of React users on SO face algorithmic issues dominantly, in which most of them are mid-reputation users (i.e., 55.77\%), followed by high and low-reputation users. 
These findings provide valuable insights for future research to assist the React community in overcoming implementation difficulties.

\textcolor{black}{Although SO remains a widely used platform for developer discussions, we acknowledge that many React developers now also seek support through other channels such as GitHub Issues, Reddit, Discord, and official documentation. Incorporating data from these alternative platforms in future research would provide a more comprehensive view of developer challenges and support-seeking behavior.
Moreover, although this study focuses on React, comparative analyses with other front-end frameworks, such as Angular or Vue, could offer valuable insights into whether the observed challenges are unique to React or common across JavaScript-based user interface libraries.}

\section{Research Methods}

In this section, we present the procedure for analyzing the React-related questions posted to SO, as illustrated in~\cref{fig:studysteps}.
The novelty of this study lies in our methodological approach, which covers multiple analyses, such as keyword identification, error types classification, and user reputation-based error types within the SO community. These analyses aim to comprehensively understand the React ecosystem, shedding light on the key themes, prevalent challenges, and expertise levels of the contributors participating in React-related SO discussions.

In detail, we present the research questions, describe the procedure of data collection, and provide our online appendix.

\subsection{Research Question}

In this paper, we formulate four research questions, completed with their motivations, that highlight the necessity for conducting this study and the approaches to conduct the investigation.

\subsection*{\textbf{\RqOne}}

\textit{Motivation:} The purpose of this RQ is to identify the most often occurring keywords in the React-related discussions that are posted on SO. 
% This is more than just a list of words; it is, at its core, an attempt to better understand the subjects that fall under the scope of React frameworks. 
By identifying the most commonly used keywords, we explore the important topics and ideas that are discussed in the online community. Essentially, the goal of this study is to shed light on the essential elements that draw in developers and encourage discussions on React libraries in the online platform.

\textit{Approach:} To answer this research question, we applied the word cloud technique, which visually represents a list of frequently occurring words. Word cloud generates an image where the size of each word corresponds to its frequency of use~\cite{jin2017development}. To perform this analysis, we utilized the React dataset, which had undergone filtration from SOTorrent~\cite{baltes2018sotorrent}. The discussion topics for analysis were selected based on the question posts. To minimize bias, certain words such as personal pronouns (e.g., ``I,'' ``you,'' ``they,'' ``we,'' etc.) and conjunctions (e.g., ``and,'' ``then,'' ``but,'' ``or,'' etc.) were removed from the dataset~\cite{pamungkas2023investigating,firdaus2024prediction}.

\subsection*{\textbf{\RqTwo}}

\textit{Motivation:} Within the ecosystem of SO, developers frequently share challenges and problems they run into when using the React framework. To understand these issues, we categorize these challenges into distinct types based on the specific errors they represent. Through this analysis, we expect to clarify the prevalent and frequently occurring issues that lead members of the React community to ask for assistance on SO.

\textit{Approach:} 
This research question is answered through a manual analysis. To classify the error types, the first three authors defined the initial coding guide. By following the previous research~\cite{ettles2018common}, the code used to label the error types is categorized into four types, as described as follows:

\begin{itemize}
    \item \textit{Algorithmic:} This type of error stems from a flawed or inefficient approach to solving a problem. These issues typically arise at the conceptual or design level, before or during coding, and reflect incorrect problem-solving logic rather than a lack of programming fundamentals.
    
    \item \textit{Misconceptions:} 
    This category captures errors that reveal a fundamental misunderstanding of core programming concepts, language syntax, or APIs. These mistakes indicate gaps in basic knowledge and often result in logically invalid code or incorrect assumptions about how code should behave.
    
    \item \textit{Misinterpretation:} These errors arise when a programmer misunderstands the problem requirements, misreads documentation, or incorrectly interprets how a library, tool, or function is expected to work. The underlying programming knowledge may be sound, but the error arises due to incorrect assumptions or context.
    
    \item \textit{Others:} This category includes questions or problems other than the 3 error types (Algorithmic, Misconceptions, Misinterpretation). This will also include questions that do not seek to solve a programming error, but ask about general concepts, API's, tools, and non-error problems.
    % \item \textit{404:} If the question is inaccessible.
\end{itemize}

From the total of 447,542 filtered data related to React, we subsequently calculated the sample size using the survey calculator,~\footnote{https://www.surveymonkey.com/mp/sample-size-calculator/} with a confidence level of 95\% and an interval of 5. 
This calculation yields 384 random sample data. Similar to a prior work~\cite{nugroho2021project}, the three authors then individually annotated the first 30 questions from the representative sample using the defined labels. The Kappa scores were calculated to see the level of agreement between the three authors.\footnote{http://justusrandolph.net/kappa/} The kappa agreement from the three raters is 83.3\% or `almost perfect'~\cite{viera2005understanding}. Based on this encouraged agreement, the remaining samples were then coded by the second author.

\begin{table}
    \centering
    \caption{Classification of User Reputation based on Score}
    \label{tab:userreputationclassification}
    \resizebox{\columnwidth}{!}{%
    \begin{tabular}{|l|p{5.5cm}|}
        \hline
        Reputation & Score Interval \\
        \hline
        \textit{High} & if the users possess a reputation score exceeding 2,400. \\
        \textit{Mid} & if the users have a reputation score ranging between 400 and 2,400.  \\
        \textit{Low} & if the users have a reputation score less than 400.  \\
        \hline
    \end{tabular}%
    }
\end{table}

\subsection*{\textbf{\RqThree}}

\textit{Motivation:} This RQ investigates how different categories of SO users interact with React-related challenges. Particularly interesting are the users' reputation scores, which serve as representatives for their level of experience and engagement in the community on SO. This analysis involves the categorization of issues based on user reputation categories, as determined by the RQ$_2$. This classification procedure makes it easier to analyse how people with different levels of experience, as demonstrated by their reputations, interact with different kinds of React-related challenges. The goal is to learn more about the dynamic interactions between users with varying levels of expertise and the particular problem domains that draw them in and encourage them to participate in discussions.

% This RQ explores how various categories of SO users interact with React-related challenges. Of specific interest are the reputation levels of users, which indicate their experience and active involvement within the community. This facet of the study entails the categorization of issues, sorted according to the RQ$_2$, into distinct groups of users based on their reputation. This categorization process facilitates an examination of how users of varying expertise, as indicated by their reputation, engage with diverse classes of React-related problems. The overarching objective here is to gain insight into the dynamic interplay between users possessing differing levels of proficiency and the specific problem domains that captivate their attention and drive their participation in discussions. 

\textit{Approach:} To answer this question, we used the same sample data that had been manually classified according to the types of errors in RQ$_2$. We then categorized this random sample data based on the user reputation categories~\cite{movshovitz2013analysis}, as described in~\autoref{tab:userreputationclassification}.

\subsection{Data Collection}

In this study, we conducted an empirical analysis of React-related question posts shared on SO. We employed exploratory data analysis to investigate keywords, question error types, and the distribution of error types based on user reputations. To initiate the study, we collected the data from the publicly available SO data dump accessible on SOTorrent~\cite{sotorrent2021baltes}.

\textcolor{black}{The dataset used in this study covers the period from January 2012 to December 2021, covering key development phases of the React framework, including the introduction of essential features such as hooks, Redux integration, and component modularization. Although the data does not capture the most recent years, it reflects stable trends in foundational developer struggles. Future research should aim to extend this dataset to include recent posts in order to analyse emerging tools and evolving practices within the React ecosystem.}

As presented in~\autoref{tab:datacollection}, the data collection in this study was performed in four steps, as follows:

\textit{Step 1:} To carry out the study, we initially extracted 22,087,195 questions from SOTorrent, a publicly available dataset derived from the official SO data dump~\cite{baltes2019sotorrent}. 

\begin{table}
    \caption{Steps of Data Collection from SOTorrent}
    \label{tab:datacollection}
    \centering
    \resizebox{\columnwidth}{!}{%
    \begin{tabular}{|p{5cm}|r|}
        \hline
        Step & {\# SO Question Posts} \\ 
        \hline
        \textit{Step 1:} Initial raw data collection from SOTorrent & 22,087,195 \\
        \textit{Step 2:} Extraction of React-related questions using 26 specified tags & 568,419 \\
        \textit{Step 3:} Duplications removal & 447,542 \\
        \textit{Step 4:} Data sampling for manual analyses & 384 \\
        \hline
    \end{tabular}%
    }
\end{table}

\textit{Step 2:} To identify React-related questions on SO, we established 26 tags containing the specific term ``React'', as described in~\autoref{tab:freqofpostofeachtag}. Through the application of these 26 specified tags for filtration purposes, we obtained a total of 568,419 React-related question posts.

\begin{table}
    \caption{The Specified React-related Tags and the Number of Their related Question Posts}
    \label{tab:freqofpostofeachtag}
    \centering
    \resizebox{\columnwidth}{!}{%
    \begin{tabular}{|c|l|r||c|l|r|}
        \hline
        No & Tag & \# Posts & No & Tag & \# Posts   \\
        \hline
        1 & React & 348,288 & 14 & react-testing-library & 2,085\\
        2 & react-native & 110,093 & 15 & react-jsx & 1,668 \\
        3 & react-redux & 20,642 & 16 & react-select & 1,618 \\
        4 & react-router & 20,274 & {17} & react-dom & 898 \\
        5 & react-hooks & 18,379 & {18} & React-flux & 820 \\
        6 & azure-active-directory & 13,874 & {19} & create-react-app & 486 \\
        7 & react-navigation & 7,818 & {20} & React.net & 169 \\
        8 & react-native-android & 7,208 & {21} & React-testutils & 89 \\
        9 & react-bootstrap & 3,658 & {22} & konvajs-React & 79 \\
        10 & react-native-ios & 3,395 & {23} & React-native & 53  \\
        11 & react-props & 2,394 & {24} & video-React & 20 \\
        12 & react-native-flatlist & 2,219 & {25} & React-popup & 4 \\
        13 & react-apollo & 2,186 & {26} & applicationinsights-react-js & 2 \\
        \hline
    \end{tabular}%
    }
\end{table}

\textit{Step 3:} To address duplicates in the filtered raw data and minimize potential bias in our analyses, we conducted a duplication removal process, resulting in 447,542 unique SO posts. From this dataset, we applied the exploratory data analysis to investigate the keywords (RQ$_1$).

\textit{Step 4:} To gain insights into React-related error types (RQ$_2$) and the users' reputations based on these error types (RQ$_3$), we employed a manual analysis. This is conducted to better assess the conceptual nature of errors by considering extended contexts to define the problem descriptions~\cite{mccall2019new}. To ensure statistical representation, we followed previous studies~\cite{islam2021network, nugroho2021project} by counting the sample size from the 447,542 React-related questions using a survey calculator with a confidence level of 95\% and an interval of 5. This calculation yielded 384 samples.

\subsection{Online Appendix}
\label{sec:appendix}
To facilitate replication, we provide an online appendix accessible at \url{https://zenodo.org/record/6420715\#.Yk\_TwWhBxPY}. This appendix contains two data files utilized in this study: (1) the dataset comprising 447,542 React-related SO question posts, and (2) the set of 384 representative samples resulting from our manual analyses.

\begin{figure}
    \centering
    \includegraphics[width=\columnwidth]{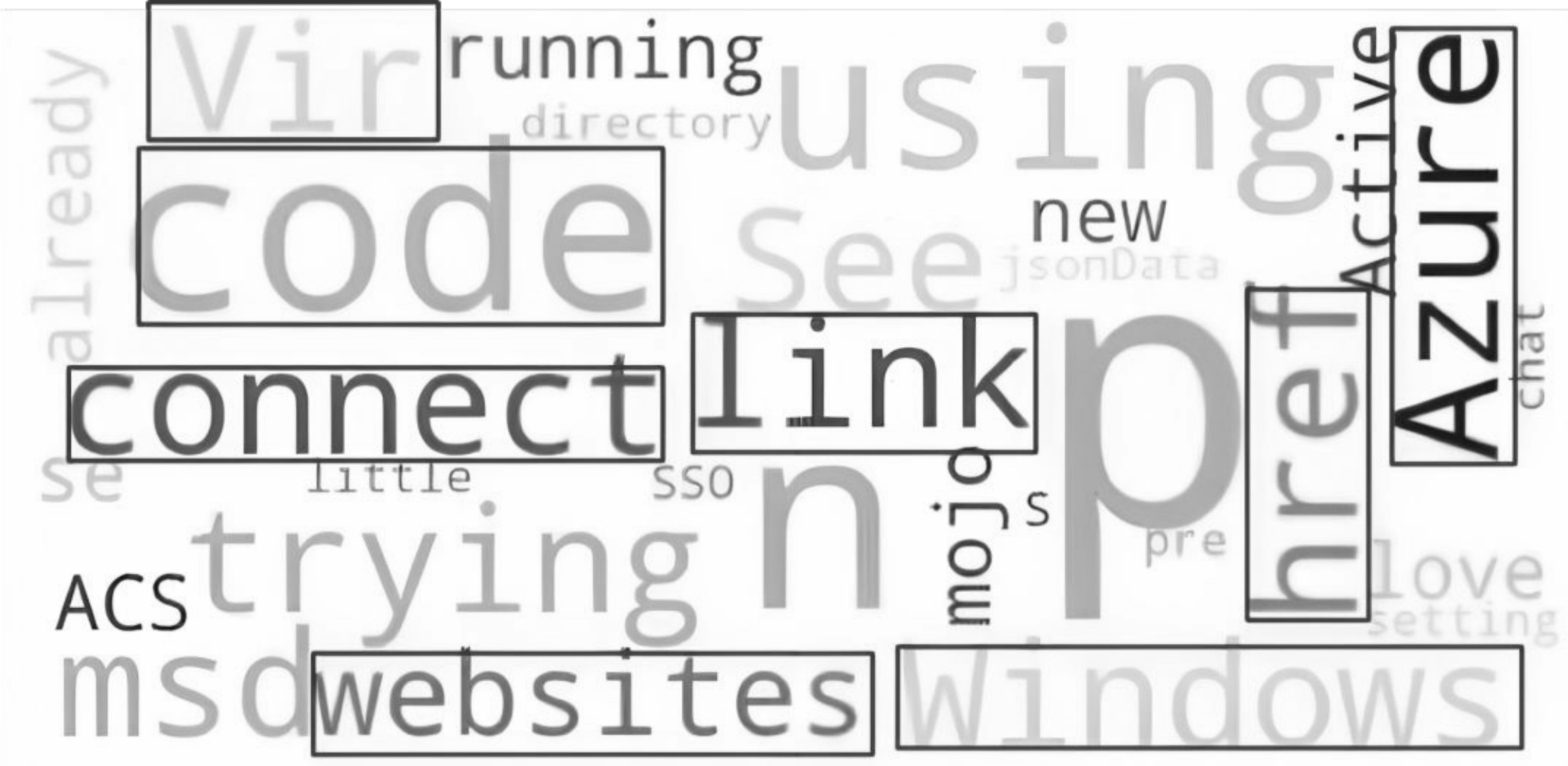}
    \caption{The most common keywords appear in React-related discussions shared on SO}
    \label{fig:keywords}
\end{figure}

\section{Results and Discussions}
\label{sec:Results}

In this section, we present the results with their discussions of each research question.

\subsection{\RqOne}

\cref{fig:keywords} illustrates the results of our analysis, which includes the top eight most important keywords extracted from what developers discussed about React, as described in~\autoref{tab:commonkeywords}.

\begin{table}
    \centering
    \caption{Top 8 Important Keywords Found in React-related Discussions}
    \label{tab:commonkeywords}
    \resizebox{\columnwidth}{!}{%
    \begin{tabular}{|l|p{6.5cm}|}
        \hline
        Keywords & Description \\
        \hline
        \textit{Code} & A term used to describe text written using a specific programming language or protocol. \\
        \textit{Link} & An interface element that establishes a connection with a target or destination.    \\
        \textit{Vir} & A tree view or structure designed for virtualization in React applications.  \\
        \textit{Href} & An abbreviation for hypertext reference, used to specify the intended web address.  \\
        \textit{Connect} & A link between React components and the Redux Store. \\
        \textit{Azure} & A Microsoft's cloud computing storage platform.    \\
        \textit{Windows} & An operating system developed by Microsoft.  \\
        \textit{Website} & A collection of web pages located within a domain or subdomain on the World Wide Web.    \\
        \hline
    \end{tabular}%
    }
\end{table}

The keywords indicate the more common topics identified by developers when discussing React, combining practical components and connections beyond React alone. Code and Link are two keywords that point to implementation and navigation, which makes sense given that developers are developing React applications. Moreover, Vir most likely refers to virtualization or virtual DOM (Document Object Model) structures, reflecting the discussions about decisions developers are making to optimize performance while working with large numbers of UI (User Interface) trees in the applications.

The keywords Connect, Azure, and Windows highlight the integration component in a broader sense, where Connect represents the popularity of state management tools like Redux in workflows where developers are used to frequently discussing or developing connections between UI components and centralized data stores. Azure and Windows both illustrate that React applications are developed and/or deployed frequently within infrastructure in enterprises, making the cross-platform usability a common discussion among professionals. 

These keywords shed light on some of the prevailing themes, tools, and environments of practical use that developers are encountering when utilizing the React framework as a front-end in working with those tools and environments.
However, although it is unclear if these eight keywords capture all main topics, they play a significant role in React-related discussions. React developers may find useful insights and solutions related to these keywords in SO threads.

\subsection{\RqTwo}

\begin{figure}
    \centering
    \begin{tikzpicture}
      \pie[
        /tikz/every pin/.style={align=center},
        /tikz/every pin/.append style={font=\footnotesize}, % Adjust the default font size for all pins
        every pin/.append style={align=center},
        sum=auto,
        after number=\%,
        radius=1.5,
        cloud,
        text=pin,
        rotate=0,
        color={black!50,black!40,black!30,black!20,black!10} % Adjust the shades of gray
      ]{
        47.92/{\color{black} Algorithmic},
        20.05/{\color{black} Misconception},
        13.28/{\color{black} Misinterpretation},
        18.75/{\color{black} Others}
      }
    \end{tikzpicture}
    \caption{The frequency of error types in our data sample of questions}
    \label{fig:errorclassification}
\end{figure}
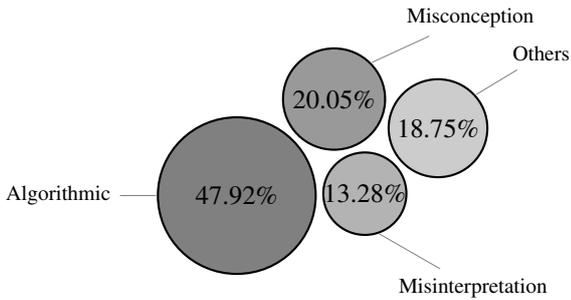

\cref{fig:errorclassification} shows that the `algorithmic' type of errors are the most frequently encountered issues related to React shared on SO. They constitute 47.92\% of the total sample of questions. The second most common error type is a `misconception,' accounting for 20.05\%, followed by a `misinterpretation' with 13.28\% of the questions. Meanwhile, 18.75\% of the React-related questions are categorized as `other', since those questions do not fit into the three defined types.

This finding indicates that React-related questions shared on SO are largely dominated by algorithmic errors, where programmers attempt to solve problems. This type of error is conceptual and may not necessarily reflect a lack of fundamental programming knowledge. This finding is well aligned with our previous work~\cite{islam2023empirical}. React users also frequently face misconceptions arising from logical flaws in their programming skills. Furthermore, React users frequently have misinterpretation problems, which occur when they commit unintentional mistakes in programming.

The result highlights the need to understand the particular types of errors faced by React developers. It implies that, in addition to emphasizing technical skills, efforts to help this community should also create a deeper understanding of React's conceptual framework and improve problem-solving techniques. These typical problems can be identified and addressed to better prepare developers to handle the challenges of React development and to build a more knowledgeable and helpful community on discussion forums like SO.

\textcolor{black}{Given the high prevalence of algorithmic errors in React-related discussions, we recommend the development of curated tutorials, visual debugging aids, and educational content that focuses on improving problem-solving strategies in React, particularly in areas such as state management, conditional rendering, and asynchronous data flows.}

\subsection{\RqThree}

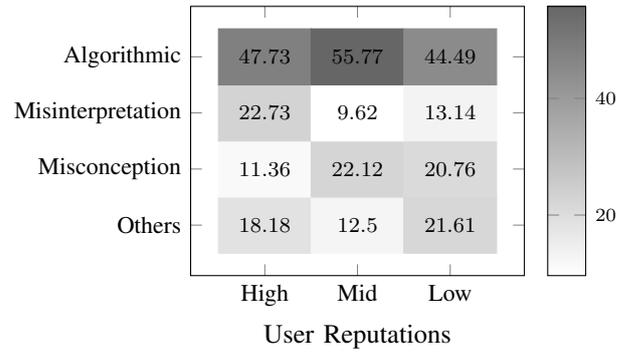
\begin{figure}
    \centering
    \begin{tikzpicture}
    \begin{axis}[
        xlabel=User Reputations,
        % ylabel=Error Types,
        xtick={1, 2, 3},
        xticklabels={High, Mid, Low},
        ytick={1, 2, 3, 4},
        yticklabels={Others, Misconception, Misinterpretation, Algorithmic},
        width=.7\columnwidth,
        yticklabel style={font=\small},
        xticklabel style={font=\small},
        colormap={graymap}{gray(0cm)=(1); gray(1cm)=(0.40)},
        colorbar,
        colorbar style={
            yticklabel style={font=\scriptsize},   % Adjust tick label font size
        },
        nodes near coords,
        every node near coord/.append style={font=\footnotesize, align=center, anchor=center}
    ]
    
    \addplot[
        matrix plot*,
        point meta=explicit,
        mesh/cols=3,
        mesh/rows=4,
    ] table [meta=C] {
        x y C
        1 1 18.18
        2 1 12.50
        3 1 21.61
        1 2 11.36
        2 2 22.12
        3 2 20.76
        1 3 22.73
        2 3 9.62
        3 3 13.14
        1 4 47.73
        2 4 55.77
        3 4 44.49
    };
    
    \end{axis}
    \end{tikzpicture}
    \caption{Frequency of error types based on user reputations}
    \label{fig:userreputation}
\end{figure}

The analysis of React-related errors across different user reputation levels on SO shows significant trends. As illustrated in~\cref{fig:userreputation}, algorithmic errors are the most prevalent across all user groups, with mid-reputation users facing them most frequently (55.77\%), followed by high (47.73\%) and low-reputation users (44.49\%). 
Misinterpretation errors are more prevalent among high-reputation users (22.73\%), while misconception-based errors are most common among mid-reputation users (22.12\%), indicating that these users may still face challenges in understanding fundamental React concepts. 
In contrast, low-reputation users contribute the highest proportion of `others' category errors (21.61\%), which could include issues unrelated to coding mistakes, such as environment setup or API integration. This suggests that while experienced users tend to misinterpret problems, mid-reputation users struggle with conceptual misunderstandings, and low-reputation users frequently face broader technical challenges beyond direct coding issues.

Comparing this result, our finding aligns with a prior work showing that algorithmic issues are dominant in JavaScript-related discussions on SO~\cite{barriere2024linear}. 
However, unlike prior studies that suggest low-reputation users contribute the most error-prone content~\cite{nugroho2021project}, our result shows that mid-reputation users face the highest proportion of algorithmic and misconception errors. 
This could be because mid-reputation users are actively experimenting with complex solutions while still refining their problem-solving skills. 
One limitation of our study is that it does not account for the time taken to resolve these errors, which could provide further insights into how efficiently different user groups overcome challenges.

The results emphasize that all user groups struggle with algorithmic issues, but mid-reputation users require targeted support to overcome misconceptions. This study is crucial in shaping community-driven learning resources, such as curated documentation and interactive tutorials adjusted to different expertise levels. Future research should explore how these error distributions evolve, considering users’ learning trajectories and their engagement with SO discussions.

\section{Conclusion}
\label{sec:conclusion}
React is a JavaScript library used to build user interfaces for single-page applications. 
While prior studies have explored general JavaScript-related challenges, a comprehensive analysis specific to React issues on SO remains limited.
We conducted an empirical investigation of React library-related question posts shared on SO using SOTorrent, a publicly accessible dataset based on the official SO data dump. 
By applying exploratory data analysis, we investigated the frequency and error classification of React-related discussions shared by developers on SO. 

Our study reveals that the topics frequently discussed by React library users on SO align closely with the eight most commonly discussed keywords. 
This finding serves as a foundation for recommending that React developers refer to SO discussions when seeking materials or solutions to address specific issues related to these keywords. 
Many React users experience algorithmic errors, which results in algorithmic error questions being a predominant topic on SO.
These insights guide overcoming inefficiencies caused by specific action biases. 

\textcolor{black}{
Although the data was collected until 2021, the identified patterns represent basic challenges that continue to affect developers even with the latest versions of React. The dataset captures core developer struggles that are likely to persist, regardless of changes in syntax or tooling.
Moreover, Large Language Model (LLM) tools such as ChatGPT and GitHub Copilot have become increasingly popular as programming assistance. While these tools may reduce developers’ reliance on Q\&A platforms for simple issues, complex logic errors and framework-specific debugging challenges still lead developers to engage in community discussions. Future work could investigate the interplay between LLM usage and developer behavior on platforms like SO.
}

\textcolor{black}{
Finally, expanding the scope of this study to other developer platforms (e.g., GitHub Discussions, Discord, Reddit) and conducting comparative analyses with other frameworks could increase the generalizability of the findings and reveal broader trends in front-end development.
}

% Based on our results, we suggest further investigation of the types of questions users asked and the relationship between user reputations and the types of questions they asked.

% \section*{Acknowledgment}
% The authors acknowledge the use of AI-assisted tools only for writing refinement in this paper. In addition, we used an LLM in this study to assess how its automatic classification compares with manual classification. However, AI was not used for content generation, data analysis, or research interpretation, and all findings result from independent work.

\balance

\end{document}